# Computer Interfaces to Organizations: Perspectives on Borg-Human Interaction Design

**Claudio S. Pinhanez**
IBM Research - Brazil
Rua Tutóia, 1157 – São Paulo, Brazil
csantosp@br.ibm.com

**ABSTRACT**
We use the term *borg* to refer to the complex organizations composed of people, machines, and processes with which users frequently interact using computer interfaces and websites. Unlike interfaces to pure machines, we contend that *borg-human interaction (BHI)* happens in a context combining the anthropomorphization of the interface, conflict with users, and dramatization of the interaction process. We believe this context requires designers to construct the *human facet* of the borg, a structure encompassing the borg's personality, social behavior, and embodied actions; and the strategies to co-create dramatic narratives with the user. To design the human facet of a borg, different concepts and models are explored and discussed, borrowing ideas from psychology, sociology, and arts. Based on those foundations, we propose six design methodologies to complement traditional computer-human interface design techniques, including play-and-freeze enactment of conflicts and the use of giant puppets as interface prototypes.

**ACM Classification Keywords**
H.5.2. User Interfaces: Theory and methods; H.5.3. Group and Organization Interfaces: Theory and method. D.2.2. Design Tools and Techniques: User Interfaces.

**INTRODUCTION**
*"The system does not allow it"* is often the frustrating answer users get when interacting with the rigidity of today's organizations. It is the epitome of the reality of our daily interaction with organizations which are complex entanglements of workers, machines, owners, buildings, rewards, systems and sub-systems, policies, associates, goals, and, too often, also other organizations.

People have to interact with such organizations to accomplish their goals in life and nowadays, more often than not, they use a computer interface (usually over the Internet) in this process. Nevertheless, we argue that most principles and practices used today in interface design are based on the assumption that users are interacting with pure machine systems, not with organizations.

The goal of this paper is to discuss how different is designing an interface to an organization vs. a machine, and to propose principles, models, and methods to address the challenges of this particular design context. The main difference, we argue, is that interfaces to organizations need to convey personality, engage in social behavior, be embodied through consistent actions, and participate in dramatic stories co-created with their users.

The main subjects of our paper are the organizations made of people, machines, and processes which surround all of us in cyberspace and with whom we are interacting constantly. We refer to them as *borgs*, paraphrasing the fictional character *Borg* of the *Star Trek* television and film series, a complex character made of billions of people, machines, and systems, but able to engage with other people displaying a remarkably consistent personality and carefully orchestrated, intentional behaviors[1].

The key characteristic of borgs is that they are large scale organizations which display complex behavior resulting of the coupling of people and the machines inside them through interactions and processes. Large scale organizations are essential components of the modern life since the Industrial Revolution and there is a vast literature studying them, including the work on sociotechnical systems (such as [27]) and on organizational theory [21].

In many cases our interaction with borgs happens through human-human interaction with the people in it (such in hospitals or schools) but more and more borg interfaces are based on computers and the Internet. Bank customers interact with their bank through its tellers and managers in a bank branch and also through ATMs and web interfaces.

---

[1] Although *forced assimilation of alien species* and *desire of universal machine perfection* were essential elements of the *Borg* character in Star Trek, those two aspects are not important or even characteristic of our use of the term *borg*.

Traditionally the CHI community has treated user interaction with pure machines and borgs almost indistinctively, as if it does not matter whether there are or there are not people and organizations behind the screen. Although the CHI practices acknowledge differences between stand-alone software and webpage interaction design (for example [15]), those differences seems to be ultimately explained as emergent properties of the Internet medium. We believe that distinguishing interaction with machines and borgs, as proposed in this paper, yields a much better understanding of why interfaces to machines and websites (which are almost always own by borgs!) are different. For instance, websites, unlike traditional stand-alone software, often raise the need of considering privacy of personal data issues. In our view, users are concerned with the personal data they are entering in a website not because they are in the web but because they know that the website is an interface to an organization which can be not only negligent with their data but also malevolent with it.

In this paper we outline and discuss what we believe are the three main distinguishing characteristics of interacting with borgs in comparison to machines. *Borg-human interaction*, or *BHI*, is, in our view, *anthropomorphized*, *conflictous*, and *dramatized*. We list those characteristics as our working hypotheses and although acknowledging that they require validation and better characterization through scientific studies, we proceed (with caution) and use them in our search for models and frameworks for understanding specific issues of borg interfaces. We then employ those frameworks to describe six design methodologies we find useful for borg-human interaction design.

**CHARACTERIZING BORG-HUMAN INTERACTION**

The underlying premise of this paper is that the presence of people inside borgs makes the interaction with and the interfaces to them fundamentally different from traditional interfaces to machines. We hypothesize here that borg-human interaction (BHI) is mainly distinct from traditional machine-human interaction because borg interaction is inherently *anthroporphized*, *conflictous*, and *dramatized*.

**Characteristic 1: BHI is Anthropomorphized**

Our daily observations of people interacting with borgs have shown us users, possibly because they know that borgs have people inside, perceive borgs as having human characteristics, treat them as (partial) human beings, and expect them to exhibit human-like behaviors. In simpler terms, unlike machines, borgs are almost always anthropomorphized up to some level by their users. People often see in large organizations human qualities such as greed, pettiness, arrogance, and evilness. We contend that users perceive similar human qualities in most machine-based interactions with borgs, such as in government websites, ATMs, search engines, webmail systems, social media sites, and, of course, airline companies.

Notice that anthropomorphization is an important characteristic of user interaction with borgs but is not exclusive to them. People attribute human characteristics to objects, places, and machines even without any trace of real connection to human beings or organizations, and change their interaction patterns accordingly, as discussed, for instance by Reeves and Nass [22].

As pointed out by Dennett [7], the complexity of most (pure) computer systems, such as machine chess players, is better dealt with by the *intentional stance*, in which the user understands the system and predicts its behavior not by knowing how it works but *"... by ascribing to the system the possession of certain information and supposing it to be directed by certain goals, and then by working out the most reasonable or appropriate action on the basis of these ascriptions and suppositions."* [7], pp. 224. However, while with most machines users have the choice adopting the intentional stance to simplify the effort of predicting the machine's behavior, we argue that in the case of borgs the actual presence of people inside them establishes anthropomorphization as the right framework for the interaction.

The important consequence of anthropomorphization is that the design and implementation of borg-human interfaces must take into account the need to provide the user with adequate representations and affordances to the perceived humanity of the borg. Today's reality is that in the majority of the cases the overall perception of the humanity of a borg is often left to be created by the user's imagination during the interaction process. To avoid this, we propose that borg-human interfaces should be structured around a coherent personality model which can be designed through specific methodologies described later.

**Characteristic 2: BHI is Conflictous**

Organizations have goals and strategies to achieve them. But more often than not, goals of people and organizations do not match, leading to some level of tension and conflict when they interact with each other. The basic consequence of this observation, in our context, is that we believe most interaction between users and borgs happens in a context of *conflict*. For instance, when a user goes into an online store, her goal is often to obtain the best of what she needs for the smallest price; in contrary, retailers "want" their customers to spend as much as possible in high-profit items.

It is interesting how most of the academic literature in interaction design tends to ignore, if not deny this conflict, which is ,in our view, a quite straightforward observation about borg-human relations. Interaction is often framed in a context of neutral dialogue, inherited from the "cold" interaction with machines; or as a collaboration process where the interface is supporting the users' goals.

A good illustration of how user-borg conflict affects interface design are the interfaces for loyalty programs such as, for example, for acquiring air tickets with miles. From a strict usability point of view, the tickets available for purchase by miles programs should be listed together with the ones that can be purchased by money, leaving to the user a clear picture of the decision he faces. The reason they are not in most airline websites is simply because the airline company goals are in conflict with the users' goals. In contrary, users' perception is often that the interface to get tickets for miles is difficult to find and difficult to use, if not intentionally slow.

Notice that although user interactions with machine-only systems are often frustrating, this tension is mostly created by the physical constraints of different types of materials, components, and sub-systems, and many times compounded with bad interface design. Though users may anthropomorphize this frustration, they do not believe that there is really bad intent (originated in values and goals) from the machines themselves, unlike in the miles case.

We argue here that in BHI the conflict is most of the time a by-product of the conflicting values and goals of users and borgs. The important question for designers is how this conflicted can be managed and, if possible, mitigated. For that, we propose to look into how human-human conflicts are dealt with, that is, through social norms and constructs, and apply those ideas to BHI design.

**Characteristic 3: BHI is Dramatized**
One way people use to make sense of their interactions with other people in life is to represent their interactions as dramatic narratives. By making ourselves heroes or victims, and rendering other people as gods or villains, we can more easily make intentions, values, and goals explicit. And by using narrative structures such as causation, succession, and counterpoint, the representation of the complex temporal patterns of our social life becomes more manageable.

Our third working hypothesis is that borg-human interaction is *dramatized* in a narrative by the user. The idea of narratives as representations or cognitive foundations for interaction is not new to CHI theory as, for example, in Laurel's work [11]. The key difference in the case of BHI is that the narrative almost always becomes dramatic: borgs are perceived as people, are in conflict with their users, and therefore can easily take the role of friends, gods, villains, or sidekicks in the narrative.

A simple example of dramatization of user interaction is often seen in the context of complaints about failures of service. Getting reparation or service recovery is in many cases described as a journey where the user is constantly facing the inability of proper contact with methods of resolution, ignorant and indifferent people, and the overall greed of the vile borg.

The appearance of a dramatic structure in BHI is a subtle characteristic which often only surfaces in more complex interactions. Nevertheless, we believe it dramatically changes the user's perception of the actions and responses of a borg's website, and therefore designers should be concerned with, and possibly design, the stories their users are creating when interacting with a borg.

**Validating the Working Hypotheses**
It is not the goal of this paper to provide empirical evidence of the validity of each of the three listed main characteristics of borg-human interaction, but instead to explore how they affect the design process in theory and practice. They are our working hypotheses for this paper, and we do see the need to explore them experimentally, through mechanisms such as structure interviews, focus groups, user surveys, and experiments such as the ones described in [22] the veracity and extent of each hypothesis.

However, in presentations and conversations we have done to professionals about those concepts, we have never found people who did not believe they were basically true. The main point of content has normally been the impact of those interaction characteristics in the design process, and whether the user's perception of the humanity of the borg (as a human being, or in conflict, or taking the role of a villain) is constant during the interaction process.

Nevertheless the theoretical need to validate and quantify the three hypotheses, which we acknowledge here, the rest of this paper is dedicated to explore how the three main characteristics of borg-human interaction, if true, can be dealt with and explored by the design process.

**MODELS OF BORG-HUMAN INTERACTION**
If borgs are perceived as humans and interact with users with human characteristics, an important set of questions arise for BHI designers. To what extent the human side of a borg has to be constructed to be perceived as an "artificial" human being, that is, how much do we need to personify the interface? Which human characteristics are more often perceived and needed by the users in borgs? When and how do users treat — and would like to treat — borgs as human beings? How to design interfaces which highlight particularly desirable human traits? How the interface can drive the drama behind the interaction process constructed by the user and better participate in it?

To address those issues, we introduce the concept of the *human facet* of a borg, which is the set and configuration of elements that create and control the perception of and the interaction with the borg's human and social characteristics. The human facet of a borg combines elements of its graphical interface, affordances, and the internal processes which together are responsible for the users' perception of the humanity of the borg. In many ways, the need to design the human facet is, as we contend in this paper, the main distinction between traditional and BHI design.

As a framework for the design of the human facet of a borg, we explore here some theoretical models of human beings which we found useful to conceive, design, and construct user interactions with borgs. To facilitate the exposition, we divide them into models of *borg personality*, *borg social behavior*, *borg embodiment,* and *borg stories*.

**Modeling the Borg Personality**

There is a vast number of proposed personality models of human beings, well beyond what could be explore in the context of this paper. As an example, we have been exploring *personality theory*, a general name for psychological models which assign archetypal categories of personality to human beings, aiming to help predict the effects of having each archetype in a given context or how each archetype normally interacts with the other archetypes.

There are two basic streams of personality archetypes. The first stream is based on the *Lexical Hypothesis* of Sir Francis Galton which proposes that personality traits are reflected into the language used by people. Based on this hypothesis, researchers in the 20[th] century refined through statistical means the most often used clusters of adjectives employed to describe people. For example, Tupes and Christal [28] and Norman [16] use five broad dimensions to describe personality traits, commonly known as *Big Five* or *OCEAN* for their initials: *Openness*, *Conscientiousness*, *Extraversion*, *Agreeableness*, and *Neuroticism* (or *Need for Stability*). *Openness* is a dimension that describes how much the person is attracted to new experiences. *Conscientiousness* describes how much the individual is able to control his or her impulses and emotions. *Extraversion* relates to how much the person is able to communicate and engage with others. *Agreeableness* describes the ability to befriend and cooperate with other people, and to be concerned with their well-being. *Neuroticism* refers to the level and need of emotional stability. Different personalities are then expressed as specific combinations of those dimensions.

The second stream of personality archetypes has its origins in Jung's *Pyschological Types* [10], which influenced, among many, the works of Briggs and Myers, who created the *Myers-Briggs Type Indicator (MBTI)* [14] which classifies individuals along four dichotomic preferences: *Extraversion vs. Intraversion (E-I)*, the preferred mode to acquire energy and motivation; *Sensing vs. iNtuition (S-N)*, determining the preferred mode to obtain information; *Thinking and Feeling (T-F)*, referring to the decision-making mechanism of choice; and *Judging vs. Perceiving (J-P)* indicating the preferred mode to relate to the world, using T-F or S-N channels, respectively. The four preferences define the 16 *MBTI types*: ESTJ, ESTP, ESFJ, and so on. The MBTI is extensively used in the industry as part of recruitment processes and in group dynamics.

More popular personality classification methods are horoscope signs, of which the most known are probably based on the Sun sign astrology (*Leo*, *Virgo*, etc.) and on the Chinese zodiac (*Rabbit*, *Monkey*, etc.). If we distanced ourselves of predictive astrology and from the simple methods to determine an individual's sign based on birthdate, we find in those schemes an interesting compendium of 12 basic human archetypes which many people easily can use as a personality classification tool.

Personality theory can make concrete and communicable to the different stakeholders in the design process the personality traits to be present in the human facet of a borg. The over-simplification of the archetype models in personality theory can be an asset on making design, engineering, and deployment needs concise and straightforward. For instance, it could be set as a design goal that *"This website should be a Virgo"*, meaning that the website should be perceived as nurturing, patient, pragmatic, loving, methodical, dedicated, and flexible — personality traits often associated with *Virgo* in the Sun sign astrology.

**Modeling the Borg Social Behavior**

Psychology has a long tradition of debating the relative importance, differences, and relationships between the personal and social aspects of the individual. Without taking sides on this century-old polemic, designers of borg-human interaction must recognize the need for interfaces to address both the personal traits of the borg (as discussed in the previous section) and the social behavior patterns exhibited by the borg in its interaction with users. Attention to social cues communicated through the interface is important not only because of the anthropomorphization of the borg but because of the conflictous relationship between users and borgs.

*Social psychology* is one of the disciplines we can draw ideas and concepts from in this context. It focuses on how social context affect human beings and how people perceive and relate to each other, therefore providing a theoretical framework to examine the interaction between the user and the borg. With the risk of some oversimplification, we can say that there are two basic currents in social psychology, coming from the psychological and sociological traditions respectively. For lack of space in this paper, we only examine basic ideas of social psychology according to the psychological stream, often associated to Kurt Lewin's work [12].

Social psychologists from this tradition divide the social phenomena into two spheres: *intrapersonal* and *interpersonal*. *Intrapersonal* phenomena of interest include the study of *attitudes*, or basic likes and dislikes; persuasion; *social cognition*, or how people collect, process, and remember information about others; *self-concept*, or how people perceive themselves; and *cognitive dissonance*, the feeling that someone's behavior and self-concept are inconsistent. For instance, cognitive dissonance increases whenever people voluntarily do activities they dislike to achieve a goal, the perception of the value of the goal is increased. So when users have to purchase miles air-tickets using a difficult interface, the overall result is to make the miles-purchased tickets more valuable than the ones bought with money, and therefore backfiring the original goal of the interface design.

Among *interpersonal* phenomena studied in social psychology which may be relevant to BHI, we can list: *social influence*, or how conformity, compliance, and obedience manifest themselves; *interpersonal attraction*, including propinquity, familiarity, similarity, physical attractiveness, and social exchange; and *interpersonal perception*, which includes issues related to the accuracy, self-other agreement, similarity, projection, assumed similarity, reciprocity, etc. For example, in interpersonal attraction, it is often true that the more someone interact with a person, the more likely she is to become emotionally engage with that person, or the *propinquity* effect. Borg-interaction designers should take the propinquity effect in account when considering the frequency of interaction between borgs and users. Frequent but short interactions may help explain the attachment of users to social networks such as *Twitter*.

Another important aspect of the social behavior is related to how emotions are used to convey and mediate social interaction between human beings. In our BHI design strategies we pay extensive attention to how users' emotions can be better understood to borgs and how borgs can use emotions to better communicate with their users. *Emotional communication theory*, which aims to understand how emotions are used in the context of interpersonal communication, is therefore an important source of models for the design of social behavior of borgs. Although research on emotions goes back to Darwin in the 19th century, the field experienced an extraordinary growth in the 1990s (see [1]). Several categorizations of emotion types have been proposed, including Ekman's [8] which proposes *happiness*, *sadness*, *fear*, *surprise*, *anger*, and *disgust* as the basic emotions. Alternatively, Russel's *circumplex model* [23] proposes a continuum representation of emotions according to two dimensions: level of *activity* (passive vs. active) and *valence* (negative vs. positive). For example, fear has high activity and some negativity, while frustration has high negativity and a moderate level of activity. Both and related theories have been incorporated before in the design of interfaces [4, 17] and we use them extensively in the design methodology is quite different.

Although studying the physiological and subjective aspects of emotion is important, the more recent research focus on how people communicate their emotions and respond to the emotional display of others during social interaction is of key interested for modeling the human facet of borgs. For instance, internal emotions can be fit to the social context, through processes such as simulation, inhibition, intensification, de-intensification, and masking (see [1], chapter3). It can be argued that for a borg to successfully manage its interaction with the user, it has to be able to adapt to the interaction context the display of its internal states that are likely to be perceived as emotions according to similar rules and methods.

**Modeling the Borg Embodiment**
Another important source of methodology for the modeling of the human facet of borgs are the many techniques used in arts and entertainment for *character embodiment*, such as the *Stanislavski's system, suspension of disbelief,* and *illusion of life*. Such concepts and techniques address how to make the human facet look real, inspire trust, and play effectively its personality, social behavior, and story role.

*Stanislavski's system* is the name associated with the methods of Konstantin Stanislavski who is often credited as the pioneer of modern acting techniques in theater. Departing from the tradition of reliance on facial expressions, excessive gesturing, and voice manipulation, Stanislavski focused on physical action: *"Acting is doing."* The best embodiments of characters do not pretend to be the characters: they act, move, and speak as the character (see Stanislavski [25]). Borg interfaces should not display carefully scripted apologies in case of failures: sorry has to be expressed with acts, such as when a voucher is given to a user to compensate for the borg's failure in delivering.

However, for borg interfaces the direct applicability of Stanislavski's and other acting methods is sometimes challenging since they were designed to create characters in human beings. An alternative body of knowledge can be used, borrowing from concepts and techniques from puppetry and movie animation, whose core issue is to vent humanity onto inanimate objects and drawings.

*Puppetry* deals almost always with the physical limitations of the puppet, with its inability to speak, to move, to have facial expressions, to have complex gestures; and nonetheless, the puppet comes alive, caring, loving, hating, and interacting with other puppets and the public. One of the contributions of puppetry to the modeling of the human facet of borgs is a very advanced understanding of which emotions can be expressed and which actions can be performed effectively in the context of limited movement and action. Hand puppets convey most of their character through talking and limited facial expression; marionettes use mostly gestures; and shadow puppetry deals with flat, black and white worlds. And yet, puppetry shows that it is surprisingly easy to make someone believe that there is an intelligent, emotional human being inside the puppet [3].

Puppetry has to take to extremes the key dramatic notion of *willing suspension of disbelief* (proposed as the center of storytelling by poet Samuel Coleridge in 1817), which require audiences to pretend to believe the puppets are real people. In BHI, the computer interface hides a real group of people behind it, and often the interface design is based on creating the illusion that the user is interacting with a cold and impersonal system. For instance, interfaces to e-commerce seem to be carefully design to hide the fact that the owner company is, in fact, an entity controlled by people with intentional, well-crafted policies to extract the largest possible amount of profit from the customer. It seems that in such interfaces to borgs, users are asked to suspend their disbelief that there are no real people behind the screen, in opposition to theater and puppetry where the goal is to make audiences suspend their disbelief that there is real life behind the characters and puppets. However, if the borg-interaction design goal is to create a true human facet to the borg using a computer-based system, we face challenges similar to the ones in puppetry, and in some ways, to the ones faced in robot-human interaction [4].

Nevertheless the goals, there is a variety of ideas and techniques from puppetry to help designers. For example, by exposing the materials and inner workings of puppets, making them move in non-realistic ways, or openly showcasing the puppeteer on the stage as in *bunraku*[2] theater, puppeteers amplify the need of belief suspension, and in the process create larger empathy between audience and characters. In the sometimes inverted suspension of disbelief of BHI, designers can expose the inner people in the borg as a way to create a stronger connection with it.

Similarly, there are lessons to be learned from movie animation, whose focus has been traditionally on how to make animated drawings convey emotions and humor. For example, Disney's animators have developed *illusion of life*, a set of 12 fundamental principles of animation [26]. For example, the *anticipation* principle states that *"[the audience] must be prepared for the next movement and expect it before it occurs. […] Before Mickey reaches to grab an object, he first raises his arms as he stares at the article, broadcasting the fact that he is going to do something with that particular object."* [26], pp. 52. In our context of BHI, we can apply anticipation by making sure that a borg's important action such as charging a credit card is clearly anticipated by actions which potentially could be stopped by the user: after the confirm button is pressed, there can be a depiction of the preparation for charging which allows one more chance for the user to change her mind (as it happens in a real store). Other fundamental principles of animation such as *staging*, *follow through and overlapping action*, *arcs*, *secondary action*, *timing*, *exaggeration*, and *appeal* may also be applied in the design of borg-human interaction animation (see [26], chapter 3).

**Modeling the Borg Stories**

While psychological models of human beings explore principles and methods from real-world human beings and their social and emotional interactions, dramatic models involve techniques to understand human behavior in the context of the dramatic stories users construct in their interaction with borgs.

A good example of a dramatic model is *character theory*, a discipline of literary and narrative studies, initially laid down by Vladimir Propp, a Russian structuralist who collected and studied hundreds of folktales and proposed that there is a common typology of narrative structures. It is based on common subsequences of 31 basic steps [19] and the identification of 8 basic roles played by what he calls *dramatis personae*, or the characters involved in a typical plot: *hero*, *villain*, *donor* (who prepares the hero for his journey), *helper*, *princess*, *princess' father*, *dispatcher* (who sends the hero off), and the *false hero/anti-hero/usurper*.

Propp claims that all folktales have similar characters and narrative structure, given and take some characters and plot steps. Similar claims can be found in the work of Joseph Campbell on mythology and mythical heroes [6], which identifies similar structures across mythologies around the world; and in Vogler's presentation of Campbell's work [29] which is extensively used in character and narrative development by the entertainment industry.

As discussed before, the interaction of a user and a borg is likely to be constructed cognitively and emotionally as a dramatic narrative where the user is often the hero. The key question for the designer is which role(s) the borg should aim to take. The borg could be the donor, the helper, or even the princess' father (the gatekeeper to the user's goals), although, in many times, it may become the villain.

Character theory has some of the concepts necessary to understand not only how to construct the human facet of the borg as a whole but also to define the different roles of the interface elements of the borg in the "fairy tale" encounter with its user. It also provides designers with a structure for human interaction with the borg based on powerful, deeply engrained psychological structures built on people from their childhood.

To finalize this discussion of dramatic models it is important to point out that many of the discussed techniques for character creation and enactment aim to *maximize conflict*, which is a major engine of dramatic success in theater and entertainment. However, in the context of borgs, we may find often that the desirable human facet is the one precisely with the opposite property, that is, a human facet which minimizes the conflict with the user. In that sense, it may be necessary to repurpose the discussed dramatics models to arrive at models that are more appropriate for the design less conflictous BHIs.

**BHI DESIGN METHODOLOGY**

After having presented and discussed the main characteristics of interaction with borgs and having explored models for those characteristics, we present here some design methodologies we are developing to address those specific issues in borg-human interaction.

We firmly believe that most traditional design methods used in computer-human interaction are also applicable to BHI, since there are many interface challenges which are basically related to the communication media (the computer screen, the hyperlink structure, etc.). We implicitly assume here that the overall BHI design process also follows basic tenants and steps of a user-centered design such as, for example, the construction of user personas [20].

---

[2] *Bunraku* is a traditional Japanese puppet theater art where puppets are manipulated by a master puppeteer wearing a traditional kimono and two black-clad, masked puppeteers. Master puppeteers show their faces to the audience but focus the audience into the puppet's action by expressionlessly and fixedly staring at the puppets.

However, the methods discussed in this section exemplify in concrete terms the need of additional work to systematically expose and target the intrinsic difficulties of creating interfaces to borgs. Inspired by the discussed models from social sciences, theater, puppetry, and comics, and some of the techniques used in those fields, we describe here six activities we believe are useful in BHI design: *back-office ethnography*, *borg personality workshop*, *conflict battle*, *puppet prototyping*, *comics workshop*, and *service blueprinting*.

**Back-office Ethnography**

As a system composed of machines, people, and processes, a borg needs to be well understood for the design process to be effective. We employ the term *back-office ethnography* to refer to the process of thoroughly investigating the inner organization and structure of the borg. We are liberally using the term ethnography here, since the actual process may include a variety of techniques, including but not exclusive to ethnography. It is inspired by some of the techniques used in *contextual design* [2] but with an additional emphasis on understanding goals and rewards.

Back-office ethnography starts with collecting all kinds of material available about the borg: organizational charts, company values, sales and production information, growth plans, etc. Even when dealing with interfaces to specific areas of a borg, it is important to collect information about the whole borg which is then complemented with a more detailed study with the issues at hand. Based on the information collected, a *borg map* is created which summarizes the basic nature of the borg.

*Goals and rewards mapping* is the next important step. Through interviews and organizational documents, we try to establish which the goals and rewards are for the different people and areas of the borg. Care should be taken to map actual, not stated goals: more often than not, the goal of many organizations is not to please the customer but to maximize revenue or profit. Similarly, rewards should be focus on actual metrics and incentives which guide the behavior of people in the organization.

The third step is to gather information about the business processes on which the interface have to rest on. The main goal is to unearth the requirements and limitations of the process to create what we call *"the system" x-ray*. The term *"the system"* is used here in the sense of the often heard sentence to justify limitations of service provision, *"The system does not allow it."* One of the best ways to produce a true picture of the "the system" x-ray is to try to use anonymously the services currently provided by the borg; and to examine customer complaints.

**Borg Personality Workshop**

Having collected information about the borg structure, goals, and processes, a designer is in the position to explore better the first characteristic of borg-human interaction, anthropomorphization, for which we have been developing a methodology called the *borg persona workshop*.

In the borg persona workshop, designers, potential users, and stakeholders try to establish the main characteristics of the borg personality from the users' viewpoint. They explore individually and in group the personality traits of the borg by using the framework of the models discussed earlier. For example, a fake Myers-Briggs test may be applied to the borg, examining the preferences of the borg as the users and stakeholders perceive it. This leads to one MBTI type, whose characteristics are then discussed.

Often, participants in the workshop are likely to differ about the borg MBTI preferences which leads to the construction of multiple personalities. This is part of the process of the borg personality workshop since there may be conflicting opinions about the desired or actual personality of the borg. It may be necessary to carry the multiplicity of borg personalities throughout most of the rest of the design process to better explore the conflicts and stories each of them generate and the different kinds of issues each personality creates. In particular, different functions of the borg interface tend to elicit distinct personalities. For instance, the sales part of a website is extroverted while the complaints interface has to be more perceiving. In certain cultures which are very familiar with horoscope signs it may be more effective to match personality characteristics of the borg with the signs.

**Conflict Battle**

In parallel with the borg personality workshop, it is often useful to run traditional CHI methodologies to determine user personas. With the different personalities of the borg and multiple user personas, the stage is then set for the *conflict battle* in which participants take turns playing the role of the borg and user personas in the different planned scenarios of BHI. The first goal of the conflict battle is to clearly document as many as possible *conflict cases*, including the situation they appear, the causes of the conflict, and how they relate to the borg inner structure. It is important to associate the conflict scenarios to the elements uncovered by the back-office ethnography process, that is, the borg map, the goals and rewards mapping, and "the system" x-ray.

The second goal of the conflict battle is to create *conflict maps* which depict the social behaviors and emotions involved in the conflict scenarios. While some of the participants are acting out the scenarios as short theatrical sketches, others should take notes of the social behaviors (such as aggression, altruism, empathy) and the emotions being exhibited by users and borg using one of the emotion characterization schemes. It is often helpful to freeze action (to be continued later) to allow time to the observers to point out, discuss, evaluate, and annotate the key characteristics of the conflict and how users and borg are dealing with it. Having the observers behind a sound-proof, see-through glass may be useful to avoid the impact of their comments on the participants enacting the conflict situation.

The third goal of the workshop is to find better ways to manage conflict and create what we call *conflict mitigation charts*. After going to the process of acting one particular scenario, participants and observers should look into possible ways of solving, mitigating the conflict, or better handling it. If necessary, alternative versions of the scenarios can be played out, trying, in particular, to examine how different personalities for the borg could cope better with the difficulties of a conflict case.

**Puppet Prototyping**
Inspired by Stanislavsky's mantra that *"acting is doing"*, the goal of *puppet prototyping* is to transform the conflict mitigation charts determined by the conflict battle into concrete interface actions which can express needed social behaviors and emotions between users and borg. While in the conflict battle we allow the full range of human actions to be played through person-to-person interaction (enacting user personas and borg), in the puppet prototyping we tunnel the interaction through representations of the computer interface using a variety of methodologies.

Many of the traditional methodologies used in interface design, such as paper prototyping, can be used here with the care of making sure to evaluate their performance in the light of conflicts and emotions associated to the different conflict cases. In association with those techniques, we also propose, especially in the initial stages of the puppet prototyping, the use of other methods inspired by theater such as *constricted dialoguing*. In this technique, participants are assigned the roles of user personas and borg and separated in two rooms connected through an interactive computer medium such as a SMS image-enriched chat. The goal is to find mechanisms, in text, interactive elements, and images, to convey the social behaviors and emotions in the conflict mitigation charts.

Another technique is what we call the *giant puppet workshop*, particularly suited for complex borg organizations with conflicting internal goals and rewards. Participants are asked to create a giant puppet, manipulated by multiple puppeteers, which has to interact, mechanically-like, with the user personas. Materials such as cardboard, colored paper, wire, glue, and recyclable elements are provided. Puppet handlers should have as much as possible goals and rewards similar to actual roles and jobs in the organization, according to the goals and rewards mapping and the overall behavior should also be constrained by the issues detected by "the system" x-ray.

The giant puppet workshop is also a technique to expose the impact of a borg with multiple personalities in the overall interaction process. In general it is not good to produce an interface to borgs with multiple-personality disorders, so the techniques described in this subsection are extremely helpful in the selection of a basic personality for the borg.

**Comics Workshop**
Having found mechanisms of communicating emotions and social behavior between users and borg and possibly the main personality characteristics of the borg, the next important step is to understand and design the stories and narratives user personas and borg will produce together.

One technique that can be employed here is what we call the *comics workshop*. It is an enriched version of the traditional storyboard technique used in interface design where participants explicit the inner thoughts of the user and the people inside the borg, the story roles (in Propp's or Campbell's sense) they play, and the overall story structure. For each interaction scenario, especially those rich in conflict, designers and participants produce a comics story showing the visual elements of the interaction, the emotional reactions of the user, and balloons with the thinking and strategy of the user, depicting, when necessary, his perception of what the borg is doing and trying to accomplish. The comics story also includes, when appropriate, the people inside the borg and what they are doing, thinking, and getting as rewards. A comics-like representation of the borg based on the giant borg puppet can also be used to convey the overall elements of the borg.

The comics stories produced in the workshop should then be analyzed in terms of character consistency, clarity, enjoyment, and conflict resolution. During this phase, elements from movie animation such as anticipation, staging, timing, and exaggeration can be tried to enhance the believability and consistency of the borg interface.

**Service Blueprinting**
The comics stories developed in the comics workshop should contain key elements of the interaction between users and borgs, but they should not be used to detail the inner workings of the borg to the point that the understanding and evaluation of the overall journey of the user is comprised.

To work out the practical details of the interaction stories of the user, a technique borrowed from service design called *service blueprinting* [24] can be used. A *service blueprint* is a representation for the user interaction with a service system (in our case, the borg) which uses horizontal tracks, or swim lanes, to represent the actions and decisions of each person and element in the system through time. The user takes the uppermost lane, which is separated from the borg lanes by the *interaction line*. From top to bottom different parts and roles of people in the system are represented, starting with the ones in direct contact with the user. All elements in the system which are never in contact with the user are grouped at the bottommost lanes, separated by the *visibility line*. Service blueprinting has been extensively used in service design and management due to its ability to highlight the influence of operational constraints, especially bellow the visibility line, in the overall user experience.

Using the comics story as a reference, a service blueprint of the borg interaction with the user in each scenario can be produced to detail the actions and decisions of each element of the borg. The service blueprints should then be analyzed in different ways. Looking vertically, information and synchronization needs of each element should be scrutinized to make sure that all their actions happen as needed. A horizontal analysis allows a good understanding of delays and waits which may affect the user experience and the borg performance and perception. Also, by considering the multiple strips from different service blueprints of a certain role or system in the borg, the complexity of its particular operation is highlighted. Here it is important to check what is expected from a role in comparison with its goals and rewards as listed in the goals and rewards mapping, and what an element of the borg should do in comparison with "the system" x-ray. Often, this analysis leads to the detection of problems which may require iterative redesign of the methodologies described.

Service blueprints, together with the other documents created in the design process, are then used to guide the development, deployment, and testing of the human facet of the borg.

**BHI Testing and Evaluation**
Before concluding our discussion of borg-human interface design, it is necessary to point out that all the effort on trying to design borg personalities, social behaviors, emotions, actions, and stories is useless if adequate and adapted interface testing and evaluations are not performed as part of the design process. However, a detailed discussion about methods to test borg-human interfaces, due to its complexity, is beyond the scope of this paper.

Nevertheless, some key issues should be mentioned. First, as in computer-human interfaces, user testing is fundamental in all stages of the design process. But in addition to traditional user testing, it is necessary to evaluate how successful the interface is in creating the essential characteristics of the borg such as the desired borg personality.

Besides verifying whether the users' is able to achieve their goals through the interface, it is also important to evaluate whether the borg's goals are also adequately met. Also, important conflict situations should be constructed and tested with real users to evaluate whether the interface is successful in managing and resolving them. Similarly, the testing process should verify whether the borg interaction stories determined in the design process actually play out as expected with real users, whether the users perceive the borg and the users roles as planned, and whether the intended story outcomes actually happen.

**RESISTANCE IS FUTILE**
The key proposition of this paper is that more often than not the interface design process is done in the context of interaction not with a pure machine system but with a complex organization of people, machines, and processes we call borgs. We argue here that designing borg-human interaction is challenged by three key differentiating characteristics of borgs to pure machines and computers: anthropomorphization, user conflict, and dramatization. We have provided frameworks to understand and model these characteristics, focusing on the borg's personality, social behavior, and embodiment, and the stories borgs co-create with their users. We have also proposed six specific design methodologies which address some key aspects and characteristics of borg-human interaction.

But do designers need really to go through all this trouble to create a good website for an organization? The "assimilation" of a borg paradigm for interface design and the adoption of BHI design methodologies is, of course, a designer's choice. In our view, the implication of creating interfaces to borgs as if they were just machines is leaving solely to the users the task of creating the human representation of the borg which, we believe, is likely to happen anyway. We have seen many cases where designing borg-human interface as if it were an interface to a machine system lead the interface to play the evil, "the system does not allow it" part of the borg.

In this paper we intentionally left out considerations and techniques for dealing with some scenarios of interacting with borgs which involve direct human-to-human contact with human beings inside the borg. Many borgs have customer care centers, physical or virtual, where the borg interfaces with its users through people. In those interactions, the personality and social behavior of the borg is also played out and borg stories are constructed. Although not addressed here, designing such human-borg interactions is essential and here some techniques from service design [9, 13, 18] should also be considered.

At the same time, the complexity of borgs keeps increasing. The spread of cheap, connected, and intelligent sensors in what often is called the age of the *Internet of Things,* coupled with the *Big Data* analytics is likely to create even more complex processes and behaviors in organizations, arguably making borgs more machine-like. In parallel, the increasing use of social media as source and media of interaction with users surges the need of coherent social behaviors and personality consistency. Borgs are becoming "borger" and, therefore, the need of better framework for designing their interaction with human beings is pressing.

The ultimate goal of this paper is to trigger an expansion of CHI into BHI. We do not deny here the importance and validity of the most interacting design theories and practices. However, we believe it is important to recognize that in the current world, computer interfaces have fundamentally changed their nature from users interacting with computers to interaction with complex organizations of people. We believe there is a brave new world of scientific opportunities to study, measure, and evaluate those interactions, and to develop and expand design to incorporate the defining characteristics of BHI.

One of the key challenges we have yet not successfully addressed is to understand what could be *sketches* of the human facet of a borg. We agree with Buxton [5] that sketching is an essential activity of the design process but traditional methods of sketching such as drawing and paper prototyping are limited, most likely inappropriate, when considering the complexity of creating human characteristics in borg interfaces. We are still far from determining useful ways to sketch the human facet of a borg, in the sense of concrete representations which *"[...] do not specify everything and lend themselves to, and encourage, various interpretations that were not consciously integrated into them by their creator."* [5], pp. 118.

Our final goal is to make designers able to create interfaces in which users can structure their relationship with borgs reliably, recognize their personality, and engage in rich social behaviors and meaningful stories. The interface to "the system" has to allow it.